\documentclass[aps,prl,floatfix,twocolumn,showpacs,amsmath,a4paper]{revtex4}
\usepackage{graphicx}
\usepackage{amsmath}
\usepackage{amssymb}
\usepackage{hyperref}
\usepackage{color}
\usepackage{oldgerm}
\usepackage{soul}
\DeclareMathOperator{\tr}{tr}
\DeclareMathOperator{\diag}{diag}
\newcommand{\comments}[1]{}
\newcommand{\bea}{\begin{eqnarray}}
\newcommand{\eea}{\end{eqnarray}}
\newcommand{\eps}{\varepsilon}

\begin{document}

\title
{Entanglement Generation is Not Necessary for Optimal Work Extraction}
\author{Karen\ V.\ Hovhannisyan,$^{1,\,*}$ Mart\'i\ Perarnau-Llobet,$^{1,\,*}$ Marcus\ Huber,$^{1,\,2,\,3}$ Antonio\ Ac\'in$^{1,\,4}$}

\affiliation{$^{1}$ICFO-Institut de Ciencies Fotoniques, Mediterranean
Technology Park, 08860 Castelldefels (Barcelona), Spain}
\affiliation{$^{2}$University of Bristol, Department of Mathematics,
Bristol, BS8 1TW, U.K.}
\affiliation{$^{3}$Departament de F\'{i}sica,
Universitat Aut\`{o}noma de Barcelona, E-08193 Bellaterra, Spain}
\affiliation{$^{4}$ICREA-Instituci\'o Catalana de Recerca i Estudis
Avan\c cats, Lluis Companys 23, 08010 Barcelona, Spain\\
{\small $^*$(These authors contributed equally to this work.)}}

\begin{abstract}

{We consider reversible work extraction from identical quantum systems.
From an ensemble of individually passive states, work can be produced only
via global unitary (and thus entangling) operations. However, we show here
that there always exists a method to extract all possible work without
creating any entanglement, at the price of generically requiring more
operations (i.e. additional time). We then study faster methods to extract
work and provide a quantitative relation between the amount of generated
multipartite entanglement and extractable work. Our results suggest a
general relation between entanglement generation and the power of work
extraction.}

\comments{
When the state of a locally thermal ensemble is factorized, the global state is also thermal and, therefore,
does not allow for work extraction. Therefore, in such a system work can be stored only in correlations. On
the other hand, entanglement is known to be a vital resource for many important tasks in physics. Now, is
this quantum resource relevant in quantum thermodynamics? To explore that, we here study how classical
and quantum correlations compare in terms of work-storing capacity. We show that for small ensembles
entangled states do indeed enable for more work than classically correlated states. However, as the size
of the ensemble grows, the work-storing capacities of classical and quantum correlations converge to the
thermodynamic bound given by the free energy difference. Also, since low-entropy global states of large
ensembles are not stable (if achievable at all), we also consider the case of macroscopic global entropy
and construct diagonal states that can store work asymptotically close to the thermodynamic bound.
}

\end{abstract}

\pacs{05.30.-d, 03.67.Bg, 84.60.-h, 03.67.Mn}


\maketitle

Energy storage and its subsequent extraction has always been a central
topic of thermodynamics, due to its obvious fundamental and practical
importance. Entanglement is a key feature of quantum mechanical systems.
It is one of the cornerstones of quantum information theory and it proved
to be important in different physical phenomena as, to name a couple,
quantum phase transitions \cite{nielsen,amico} or fractional quantum Hall
effect \cite{prange}.

In quantum thermodynamics, entanglement is also connected to work
extraction from multipartite systems. Indeed, global (and thus
entangling) unitary operations are capable of extracting more work than
local operations from a set of quantum systems (see, for instance,
\cite{alicki}). However, that an entangling operation is needed for a
process, does not imply that entanglement is generated during its
execution. The scope of this work is to clarify these connections. We
first show that optimal work extraction can be achieved without any
entanglement generation. The corresponding dynamics is slow, in the sense
that it requires many different operations. We then consider faster
methods and establish a link between entanglement generation and
extractable work. Our results point to a connection between entanglement
generation and work power.

A classical result on work extraction is the no-go theorem called
``\emph{Thomson's formulation of the second law of thermodynamics}'': no
work can be extracted from a thermal state via cyclic hamiltonian
processes \cite{ll5}. It is extended to finite quantum systems by
introducing the concept of {\it passive} states, those not capable of
giving out energy during any cyclic hamiltonian process \cite{pusz,
lenard}. Thermal states are passive, while not all passive states are
thermal. Indeed, several copies of states of the latter type may contain
population inversions, allowing for some work extraction. On the other
hand, thermal states do not share this ability of "activation" since any
combination (tensor product) of thermal states is a thermal, thus passive,
state. The converse is also true: if any combination of a passive state
is passive then it is a thermal state. One can, however, get some work
out of a set of locally thermal states by initially correlating them, the
microcanonical state being a prominent example \cite{armen,parrondo,
jarzynski,akepl}.

We study the problem of work extraction from $N$ noninteracting finite-
level systems. For simplicity and with no loss of generality we assume
all systems to be identical with $d$-level hamiltonian
$H=\sum_{k=1}^d \eps_k |k\rangle\langle k|$, $\eps_1\leq...\leq\eps_d$.
The hamiltonian of the ensemble is then $h_0 = H \otimes \cdot \cdot \cdot
\otimes 1 + ... + 1 \otimes \cdot \cdot \cdot \otimes H$. We take the
eigenbasis of $h_0$ as the standard basis.

Typically one wishes to store (even if not for long time) the energy before
extracting it. For that, one prepares the ensemble in an active state which
is diagonal in the total hamiltonian so that it remains unchanged:
\bea \label{state}
\Omega=\diag\left( P_1,\,...,\,P_{d^N} \right)
\eea
with all $P_\mu\geq0$ and $\sum_{\mu}P_\mu=1$. Note that
the initial state is separable, as it is diagonal in the product energy
eigenbasis, but may display classical correlations.

Now, to extract work, at the moment $t=0$ external control fields with a
time-dependent potential $V(t)$ are turned on. In the presence of these
fields the ensemble|made up of $N$ systems|undergoes a unitary evolution
driven by the hamiltonian $h(t)=h_0+V(t)$. Then, at the moment $t=\tau$,
external fields are turned off: $V(\tau)=0$; providing, thus, the cyclicity
of the process and returning the systems their original structure. The overall
unitary evolution operator is given by the time-ordered exponent $U(\tau)=
\overrightarrow\exp\left(-i\int_0^\tau d t\left[h_0 + V(t)\right]\right)$,
rendering the final state to be:
\bea \label{xuy}
\Omega(\tau)=U(\tau)\Omega U^\dagger(\tau).
\eea
The work extracted during the process is given by the difference of the
initial and final energies of the system:
\bea \label{w}
W=\tr\left(\Omega h_0\right) - \tr\left(\Omega(\tau) h_0\right).
\eea
As proven in \cite{armen}, the unitary operator minimizing ${\rm tr}\left(
\Omega(\tau) h_0\right)$ (hence maximizing $W$) is the one that results in
a permutation of the elements of $\Omega$ so that the largest element of
$\Omega$ is matched with the smallest one of $h_0$, the one-but-largest
with the one-but-smallest, etc. This shows that the state is passive iff
its elements are reversely ordered with respect to the hamiltonian $h_0$.

Now note that after maximal work extraction the final state of the ensemble
is separable because is diagonal in the eigenbasis of the non-interacting
hamiltonian $h_0$, as the initial state. On the other hand, permutation
operations are not local and can have maximal entangling power \cite{ent_perm},
so the following natural questions arise: does the state of the system get
entangled during the process? If yes, how entangled does it become? Is there
any way to bypass the entanglement creation, so that the state remains
classically correlated all the time?

We start by answering the last question and provide a protocol that attains
maximal work extraction with no entanglement. Then, we consider faster
protocols and provide lower bounds to the entanglement they generate. These
bounds are applied to two relevant scenarios, namely maximal work extraction
from the microcanonical bath \cite{armen,parrondo,jarzynski,akepl} and from
copies of passive states \cite{pusz,lenard}.

{\it Bypassing entanglement. Indirect paths.}|To extract maximal work from
a state of the form (\ref{state}) we need to reorder its entries accordingly.
This reordering can be done in elementary steps of transpositions. E.g.,
$\Omega$ may be such that the population of the lowest energy level ($P_1$)
is not its maximal element, {namely $P_{\nu\neq1}$}. Then one needs to
transpose $P_1$ with $P_{\nu}$, etc. After some number of such steps the
state will be ordered properly, yielding maximal work.

The transposition of the population of any energy level by the population
of some other level can be done without creating entanglement in the meantime.
Indeed, suppose we need to transpose $P_\alpha$ by $P_\beta$. The eigenstates
of $h_0$ corresponding to them are $|\alpha\rangle=|i^\alpha_1\,i^\alpha_2...
i^\alpha_N\rangle$ and $|\beta\rangle=|i^\beta_1\,i^\beta_2...i^\beta_N\rangle$,
respectively. We then divide this action in $2N-1$ transposition steps. First
\bea \label{long}
|i^\alpha_1 i^\alpha_2...i^\alpha_N\rangle\rightleftarrows|i^\beta_1 i^\alpha_2
...i^\alpha_N\rangle\rightleftarrows|i^\beta_1 i^\beta_2...i^\alpha_N\rangle ...
\rightleftarrows|i^\beta_1 i^\beta_2...i^\beta_N\rangle~~~
\eea
and the $N-1$ steps back from $|i^\beta_1 i^\beta_2...i^\beta_{N-1} i^\alpha_N
\rangle$ to $|\alpha \rangle$. On each step one only exchanges the populations
between states involved in it. Here we stress that all steps in (\ref{long})
involve only the corresponding states, while the populations of the other
states are kept unchanged. Thus, one cannot perform any of these steps with
a local unitary operation. Here we assumed the number of positions where the
inputs of strings $\{i^\alpha_k\}_{k=1}^N$ and $\{i^\beta_k\}_{k=1}^N$ are
different, $n^{\alpha\beta}$, to be equal to $N$. In general, $n^{\alpha\beta}$
can be $<N$ making the chain (\ref{long}) shorter. This generalization can
straightforwardly be done throughout the article.

The unitary operator that transposes the populations of two basis states, say $|\alpha
\rangle$ and $|\beta\rangle$, reads as
\bea \label{trans} \label{Utransposition}
U^{\alpha\beta}=\sum_{\mu\neq\alpha,\beta}|\mu\rangle\langle\mu|+
|\alpha\rangle\langle\beta|+|\beta\rangle\langle\alpha|.
\eea
If the control potential $V(t)$ generating $U^{\alpha\beta}$ couples only
to $|\alpha\rangle$ and $|\beta\rangle$, the evolution operator at some
intermediate moment $t$ of the process is
\bea \label{aranq}
U^{\alpha\beta}(t) = \sum_{\mu\neq\alpha,\beta}|\mu\rangle\langle\mu|+u^{\alpha\beta}(t),~
\eea
where $u^{\alpha\beta}(t)$ lives in the linear span of $|\alpha\rangle$ and
$|\beta\rangle$ and is unitary. It depends on $t$ and the concrete form of $V(t)$.

Now for, e.g., the first step in (\ref{long}) we need to perform the transposition
unitary $U^{\alpha\alpha^\prime}$ between $|\alpha\rangle=|i^\alpha_1 i^\alpha_2...
i^\alpha_N\rangle$ and $|\alpha^\prime\rangle=|i^\beta_1 i^\alpha_2...i^\alpha_N\rangle$.
According to (\ref{aranq}) the state of the ensemble at an intermediate moment $t$ is
\bea \nonumber
\Omega (t) = U^{\alpha\alpha^\prime}(t)\,\Omega\, U^{\alpha\alpha^\prime\dagger}
(t) \\ \label{sep_form} && \hspace{-42mm} = \hspace{-1mm} \left(P_\alpha\hspace{-0.2mm} +P_{\alpha^\prime}\right)\hspace{-0.2mm}\rho_1(t)\hspace{-0.3mm}\otimes\hspace{-0.3mm}
|i_2^\alpha...i_N^\alpha\rangle\langle i_2^\alpha...i_N^\alpha|\hspace{-0.2mm}+\hspace{-2mm}
\sum_{\mu\neq\alpha,\alpha^\prime}P_\mu |\mu\rangle\langle \mu|.~~~~~~\hspace{-2mm}
\eea
Quite straightforwardly, $\rho_1(t)\geq0$ and $\tr [\rho_1(t)]=1$, so (\ref{sep_form})
means that the state of the ensemble is separable during the whole process of population
exchange between $|\alpha\rangle$ and $|\alpha^\prime\rangle$. Notice that although
$U^{\alpha\alpha^\prime}(t)$ is global and thus has entangling power, there exist states
which it does not entangle.

By the same reasoning, one may stay separable also during the rest
of transpositions in chain (\ref{long}). So any replacement in the
global state of the ensemble can be made without creating
entanglement between its constituents, which proves that one can
extract maximal work from the ensemble and stay separable during
the whole process.

Finally we note that the results of this section also hold for a
generic product initial state $\otimes_i \rho_i$, where each
$\rho_i$ may have quantum coherences, that is, off-diagonal elements
in the energy eigenbasis. One first extracts maximal work from
individual systems via local, thus non-entangling, unitary operations, and
then applies the protocol described above to the resulting
diagonal state.

The previous non-entangling protocol requires $2N-1$ global
operations in order to perform the desired exchange of the
populations $|\alpha\rangle$ and $|\beta\rangle$. However,
this exchange can be performed in one step by the unitary operator
(\ref{trans}). We term such evolutions by direct paths. Now, the
natural question is whether entanglement is generated by these
direct paths, which allow one to extract work faster and thus get
more power.

{\it Direct paths.}|Consider the population exchange of $|\alpha\rangle$
and $|\beta \rangle$. A relevant example of a direct path would be the
time independent hamiltonian $H=\frac{\pi\hbar}{2\tau}\left(|\alpha\rangle
\langle\beta|+|\beta \rangle \langle \alpha | \right)$ which generates the
desired interchange at $t=\tau$. More generally, we will consider the
evolution of $\Omega(t)=U^{\alpha \beta}\Omega U^{\alpha\beta\dagger}$
where $U^{\alpha \beta}$ is found from (\ref{aranq}).

In order to measure the entanglement of $\Omega(t)$ in a direct path we use
a recently proposed measure of genuine multipartite entanglement for mixed
states \cite{ma,maju,mamaju} which luckily turns out computable for states
relevant to the work extraction protocol \cite{np}. The measure is essentially
a straightforward generalization of the concurrence \cite{conc} to multipartite
systems (see the Appendix for a detailed description
of the measure), and therefore has a clear operational meaning.

The measure represents an ordered string $E$ (with elements $E_1\geq\cdots\geq
E_{2^{N-1}-1}\geq0$) called entropy vector (see the Appendix for the definition),
which quantifies multipartite quantum correlations the following way: if the
last $2^{l-1}-1$ entries of $E$ are zero, then the state is $l$-separable
\cite{mamaju}. For pure states the latter notion means that the state vector can
be written as a tensor product of at most $l$ terms. Then, a density matrix is
said to be $l$-separable if it can be decomposed into pure states that are at
least $l$-separable. Two immediate corollaries of the above definitions are: (i)
the familiar fully separable (nonentangled) state is the $N$-separable state,
for which $E$ is the zero string. That is why $E_1>0$ is the necessary and
sufficient condition for the state to be entangled. (ii) The $1$-separable
state is the genuinely $N$-partite entangled state \cite{l_sep}. Therefore,
the smallest entry of $E$, $E_{2^{N-1}-1}$, measures the amount of genuinely
$N$-partite entanglement in the state.

In the Appendix we bring closed-form expressions of
lower bounds for all entries of the entropy vector. The states we deal with in
this work have only two (complex-conjugated) nondiagonal elements (\ref{state},
\ref{xuy}, \ref{aranq}). The latter fact greatly simplifies the formulas for the
mentioned lower bounds, and their maximal values (reached simultaneously) during
the transposition between $|\alpha\rangle$ and $|\beta\rangle$, $\Lambda_k$, are
as follows (see the Appendix):
\bea \label{evlb}
E_k\geq\Lambda_k=|P_{\alpha}-P_{\beta}|-2\min_{A}\sum_{a\in \Gamma^k_A}\sqrt{P_{\alpha_a}P_{\beta_a}},
\eea
where $a$ runs over all bipartitions $\gamma_a\cup\bar{\gamma}_a$ of the set
$\{1,...,N\}$; $A$ enumerates the set of all k-tuples $\Gamma^k_A$ of the index
$a$; $|\alpha_a\rangle$ is obtained from $|\alpha\rangle=|i^\alpha_1...i^\alpha_N\rangle$
by replacing $i^\alpha_k$ by $i^\beta_k$ for all $k\in\gamma_a$; analogously,
the replacement series $\{i^\beta_k \to i^\alpha_k\,:\,k\in\gamma_a\}$ takes
$|\beta\rangle=|i^\beta_1...i^\beta_N\rangle$ to $|\beta_a\rangle$; as is proven
in \cite{maju}, $E_1=\Lambda_1$ and $E_{2^{N-1}-1}=\Lambda_{2^{N-1}-1}$. (Note
that after each transposition step the entries $P_\mu$ of (\ref{state}) must
be updated before using the formula (\ref{evlb}).)

{\it Microcanonical bath.}|As a first application, consider work extraction
from the microcanonical state of width $\Delta$ around some energy
$\mathcal{E}_0$ of the total system
\bea \label{mc}
\Omega_{mc}=\frac{1}{N_\Delta}\sum_{m=1}^{N_{\Delta}}{|\mathcal{E}_0^{(m)}\rangle
\langle \mathcal{E}_0^{(m)}|}
\eea
with $|\mathcal{E}_0^{(m)}\rangle$ being states with energy in the interval
$[\mathcal{E}_0-\Delta/2,\,\mathcal{E}_0+\Delta/2]$ \cite{zhopa}. This state is active
\cite{armen,parrondo,jarzynski,akepl} but is not capable of giving macroscopic
work in thermodynamic ($N\gg1$) limit \cite{akepl}.

In this setting, work is extracted by performing $N_\Delta$ exchanges of
the populations (all equal to $0$) of the $N_\Delta$ states $|\epsilon_0^{(m)}
\rangle$ ($m=1,...,N_{\Delta}$) with the lowest energy by the populations
(all equal to $1/N_\Delta$) of the states $|\mathcal{E}_0^{(m)}\rangle$.
These exchanges do not overlap and can be done in $N_\Delta$ successive
steps. If as a first exchange we take the one between states with energy
$0$ and $\mathcal{E}_0-\Delta/2$, then if there are $n_1$ different indices
in those sates \cite{n_1}, formula (\ref{evlb})
shows that the corresponding bath particles are genuinely $n_1$-partite
entangled during the whole transposition. \cite{foot}

{\it A set of passive states.}|Another important scenario is the product of
passive-but-not-thermal states \cite{alicki,pusz,lenard}:
\bea \label{ap}
\Omega_{ap}=\sigma_p^{\otimes N}.
\eea
with $\sigma_p=\diag(p_1,...,p_d)$. 
The maximal work $W$ is delivered by the unitary operation, that minimizes the final
energy. The unitary operation preserves the whole spectrum of the state. Nevertheless, if we
require our transformation to preserve only the von Neumann entropy $S(\sigma)
=-\tr[\sigma\ln\sigma]$, the consequent minimal energy will generally be
less and will be delivered by a thermal state \cite{alicki}. So, subtracting
the latter minimal value from the initial energy will upper-bound the work $W$.
Moreover, as is shown in \cite{alicki}, such a bound can be reached
in the thermodynamic limit, i.e.,
\bea \label{al_wo}
W \overset{N\to\infty}{\longrightarrow} N\tr\left[ H(\sigma_p-\sigma_{\rm th}) \right]=N T S(\sigma_p||\sigma_{\rm th}),
\eea
where $\sigma_{\rm th}=\diag(q_1,...,q_d)$ is thermal and has the same entropy as $\sigma_p$,
and $S( \hspace{1mm}||\hspace{1mm} )$ is the conditional entropy.

Observe that for $N\gg1$ the spectrum of $\sigma_p^{\otimes N}$ has a
typical subset -- a set of almost identical eigenvalues that asymptotically
sum-up to $1$ \cite{cover}. Hence, the number of elements in the typical set is
asymptotically equal to $e^{NS(\sigma_p)}$ \cite{typic}. Because
$S(\sigma_p)=S(\sigma_{\rm th})$, the typical sets of $\sigma_p^{\otimes N}$
and $\sigma_{\rm th}^{\otimes N}$ coincide.

Since the overall probability of non-typical states is exponentially small,
upon transposing in $\sigma_p^{\otimes N}$ the populations of its typical
states with the ones of states typical in $\sigma_{\rm th}^{\otimes N}$, we
arrive at an energy approximately equal to the one of $\sigma_{\rm th}^{\otimes N}$
\cite{alicki}. These $e^{NS(\sigma_p)}$ transpositions do not overlap and
can therefore be made successively.

Consider the direct path exchange of the populations of $|\alpha\rangle$
and $|\beta \rangle$ given, respectively, by $P_\alpha=\prod_{j=1}^{N}
p_{i_j^{\alpha}}$ and $P_\beta=\prod_{j=1}^{N} p_{i_j^{\beta}}$. Here the
expressions (\ref{evlb}) are simplified to
\bea
E_k\geq \left|P_\alpha-P_\beta\right| -2k\sqrt{P_\alpha P_\beta}.~~~
\eea
So, e.g. for $P_\alpha \geq P_\beta$, the state will be at most
$l$-separable when:
\bea \label{condentt}
\frac{P_\alpha}{P_\beta}\geq 1+2\gamma+2\sqrt{\gamma+\gamma^2},
\,\,\, \gamma=2^{N-1}-2^{l}+1.~~~
\eea

Now, pick one state from the typical set of $\sigma_p^{\otimes N}$, say
$\bigotimes_{k=1}^d|k\rangle^{\otimes N p_k}$, and transpose its probability
$\prod_{k=1}^d p_k^{N p_k}$ with $\prod_{k=1}^d p_k^{N q_k}$ -- the population
of the corresponding state $\bigotimes_{k=1}^d|k\rangle^{\otimes q_k}$
from the typical set of $\sigma_{\rm th}^{\otimes N}$. Then the formula
(\ref{condentt}) will imply the following condition
\bea \label{condenttropy}
S(\sigma_{\rm th}||\sigma_p)\geq \frac{1}{N}\ln \left[1+2\gamma+2\sqrt{
\gamma+\gamma^2}\right]
\eea
that $\sigma_p$ must satisfy to be at most $l$-separable during the process. Here
$\gamma$ is the same as in (\ref{condentt}). It can be easily checked that
(\ref{condenttropy}) is the same for all typical states, so it holds for the whole
work extraction process.

Condition (\ref{condenttropy}) has a simple interpretation -- the greater
the difference between $\sigma_p$ and $\sigma_{\rm th}$, the more
entanglement we need. In the $N\gg1$ limit, the condition for entanglement
to be present is $S(\sigma_{\rm th}||\sigma_p)\geq \ln [3+2\sqrt{2}]/N$
so basically all states get entangled, while the condition for genuine
$N$-partite entanglement to appear is $S(\sigma_{\rm th}||\sigma_p)\geq
\ln [2]$ which tells that this entanglement does not have to be $N$-partite.

\begin{figure}[t]
   \includegraphics[width=0.47\textwidth]{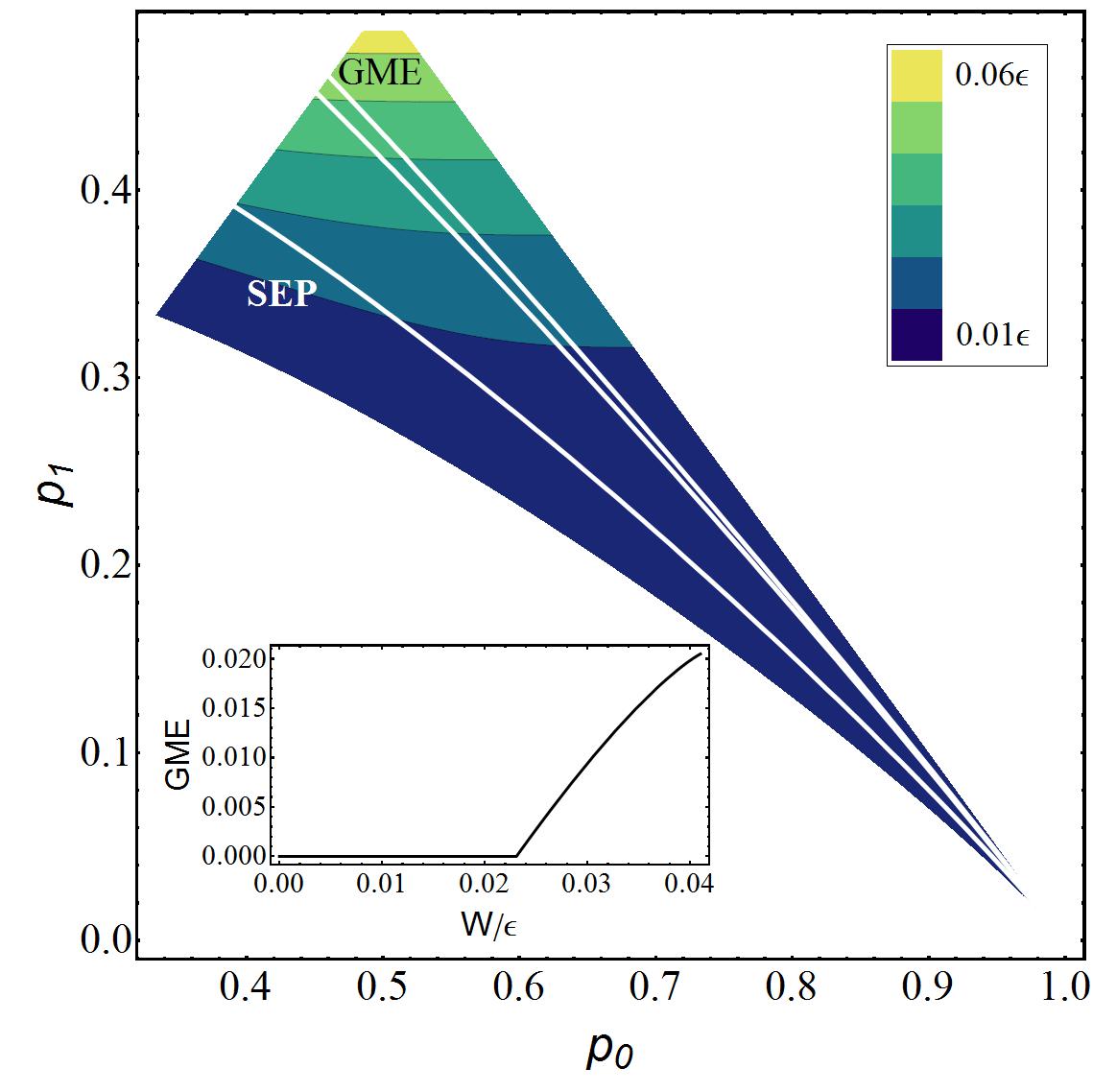}
    \caption{A contour plot of the work $W$ released by four three-level systems
    initially in the state $\otimes^4 \sigma_p$ on the direct path exchange of
    the populations of levels $|1111\rangle$ and $|0222\rangle$. The levels of
    each system are $\{0,\epsilon,\epsilon\}$. Lighter regions correspond to
    more work extraction. The white lines separate regions of $l$-separability,
    the left side (SEP) being fully separable and the rightmost region (GME) 
    genuinely multipartite entangled. Notice that there are regions where work
    is extracted without generating any entanglement. However, as $W$ increases,
    $k$-partite entanglement starts appearing with $k$ progressively larger.
    The inset illustrates the direct quantitative relation between the amount
    of genuinely 4-partite entanglement measured by ${\rm GME}=E_{2^{N-1}-1}$, and
    the extractable work $W$ in the same setting and with $p_0=0.55$.
}
\label{fig:entwork}
\end{figure}

On the other hand, the extracted work is governed by the difference of
$\sigma_p$ and $\sigma_{\rm th}$ (\ref{al_wo}). So the farther $\sigma
_{\rm th}$ is from $\sigma_p$ the more entanglement is generated and more
work is extracted. This situation is illustrated in Fig.~\ref{fig:entwork}
for four three-level systems.

{\it Entanglement and power.}|These two exemplary cases show that
entanglement is widely present during direct routes. Furthermore,
the amount of entanglement is directly connected to the amount of
work for the case of identical systems. In general this connection
exists but is less direct: work from an elementary exchange is
proportional to the difference of populations involved, and so is the
first term in (\ref{evlb}) -- the expression for generated
entanglement.

On the other hand, the amount of generated entanglement can be
reduced by combining direct and indirect paths. One simply performs
$N-l$ exchanges via indirect paths followed by a direct path producing
at most $l$-partite entanglement. Alternatively, (\ref{condenttropy})
implies that for identical systems one can arbitrarily reduce the
amount of entanglement generated by performing $K$ extra steps of direct
exchange of states $\rho_k$ ($k=1,...,K$) satisfying $S(\rho_1||\sigma_p)
<S(\rho_2||\sigma_p)<...<S(\sigma_{\rm th}||\sigma_p)$. In both cases,
if we assume that all global transpositions are equally time consuming,
reducing entanglement production comes at the expense of increasing the
time of the process.

Our analysis thus suggests that, although entanglement plays no role for
the amount of work one extracts, it may be crucial for the power -- the
larger the power output the more entanglement is created during the process.
This strictly holds for the protocols considered in this work, and it does
not depend on the choice of the entanglement measure.

{\it Conclusions \& Outlook.}|We studied the role of entanglement generation
in work extraction from an ensemble of $N$ non-interacting systems.

We introduce a protocol for maximal work extraction such that no entanglement
is created in the ensemble during the runtime. We then consider direct paths
and show that a higher amount of work extraction requires a higher level of
entanglement (see, e.g., Fig.~\ref{fig:entwork}). For the considered protocols,
the power at which work is extracted is connected to the entanglement production.

In recent proposals of resource theories for work extraction Ref.~\cite{Skr,brandao},
the latter is also studied as a permutation of diagonal elements in a density
matrix, so our results regarding the entanglement generation straightforwardly
apply there. Actually the task of population exchange is also essential in other
processes in quantum thermodynamics such as dynamical cooling of spins in NMR
physics or constructing optimal heat engines (see, e.g., \cite{sorensen,ajm,
LindenSkrzypczyk}), etc., so our results have implications also beyond the problem
of work extraction.

We thank D. Cavalcanti, A. E. Allahverdyan, J. B. Brask, A. Bendersky, J. Calsamiglia
and P. Skrzypczyk for helpful discussions. This work is supported by the ERC Starting
Grant PERCENT, and the Spanish FIS2010-14830 project and Severo Ochoa program. MH
acknowledges the MC grant "Quacocos" (N302021).

\section*{Appendix} \label{appendix}

In this Appendix we expand the explanations of the used entanglement
measure and its lower bounds in more detail. To introduce the mentioned measure,
consider an $N$-partite system in a pure state $\psi$. We define its entropy vector
$\vec{S}$ \cite{winter,maju} as a string, whose elements are the entropies of the
reduced states of all subsystems of our $N$-partite system:
\bea
\label{entropyvect}
S_k(\psi)=\sqrt{2(1-\tr(\rho_k^2))}, \,\,\, k={1,...,2^{N-1}-1},
\eea
where the index $k$ runs over all possible bipartitions
$\Gamma=\{(\gamma_k|\bar{\gamma_k})\}$ of $\{1,2,...,N\}$; and
$\rho_k$ is the corresponding reduced state: $\rho_k={\rm
Tr}_{\bar{\gamma_k}} (|\psi \rangle \langle \psi|)$.

The linear
entropy is used in (\ref{entropyvect}) for mathematical convenience. Similarly to the
standard entropy of entanglement for bipartite pure states, a non-zero (linear) entropy
of $\rho_k$ reflects the presence of entanglement. Therefore, each entry of the vector
(\ref{entropyvect}) detects entanglement in a particular bipartition.

The entropy vector  (\ref{entropyvect}) also reveals multipartite entanglement properties of the
state, in particular the $l$-separability \cite{mamaju}. A pure state is $l$-separable
if it can be written as a tensor product of at most $l$ terms, i.e., in the form $\rho_1
\otimes \rho_2 \otimes ... \otimes \rho_k$. Then, it is easy to realize that a $l$-separable state contains $2^{l-1}-1$ bipartitions where the state is separable.  Therefore, if we set
the entries of $S_k$ in non-increasing order, (i) if the last $2^{l-1}-1$ entries are
zero, then the state is $l$-separable. In particular, (ii) if the first entry is zero, then the state is entangled (i.e., at least $(N-1)$-separable). Furthermore, (iii) if the last entry is non-zero, i.e. all entries are non-zero, then the state is genuinely multipartite entangled (GME), or $1$-separable. Indeed the last entry can be used to measure GME \cite{hashemi}.

This measure is extended to mixed states $\rho$ via a convex
roof construction \cite{maju,mamaju}:
\bea \label{roof}
E_k(\rho)\equiv \inf_{\{p_i;|\psi \rangle\}} \sum_i p_i S_k(\psi_i),
\,\,\, k={1,...,2^{N-1}-1},~~~~~
\eea
where the infimum is taken over all
possible decompositions $\{p_i;|\psi \rangle\}$ of $\rho$, and
$S_k$ are the entries of the entropy vector arranged in
non-increasing order. Then, the statements (i), (ii) and (iii) above also
hold for $E_k$ ($E_k$ plays the role of $S_k$ in mixed states).

Since the convex roof construction (\ref{roof}) requires an optimization over an infinite set
and even determining whether a single entry is nonzero is, in general, NP-hard (\cite{gurvits}),
we have to find a method to reliably calculate lower bounds to these measures.
Refs.\cite{ma,wu,se,maju,mamaju} provide lower bounds on all entries of the entropy
vector in terms of the matrix elements of the state $\rho$. For a given set $\mathcal{C}$ of off-
diagonal elements of $\rho$, they read:
\comments{
a framework constructing general nonlinear entanglement witness inequalities, whose degree
of violation provides lower bounds to all entries has been developed, yielding:
}
\bea
\label{eqEks}
E_k\geq\Lambda_k (\mathcal{C})=\frac{1}{\sqrt{|\mathcal{C}|}}\sum_{(\alpha,\beta) \in \mathcal{C}} \left[|\langle
\alpha | \rho | \beta \rangle |
\right.
 \nonumber\\
\left.
-\min_{A}\sum_{a\in \Gamma^k_A} \sqrt{ \langle \alpha_a | \rho | \alpha_a \rangle \langle
\beta_a | \rho | \beta_a  \rangle}  \right],
\eea
where the index $a$ runs over all bipartitions $\gamma_a\cup\bar{\gamma}_a$ of the set
$\{1,...,N\}$; $A$ enumerates the set of all $k$-tuples $\Gamma^k_A$ of the index
$a$; $|\alpha_a\rangle$ is obtained from $|\alpha\rangle=|i^\alpha_1...i^\alpha_N\rangle$
by replacing $i^\alpha_k$ by $i^\beta_k$ for all $k\in\gamma_a$ and analogously for
$| \beta_a \rangle$.

The lower bounds (\ref{eqEks}) easily apply to the considered processes in this work.
Indeed, given the initial global diagonal state $\Omega$, consider the exchange of
populations $P_\alpha$, $P_\beta$ of the states $|\alpha \rangle$, $|\beta \rangle$
respectively; under some unitary process given by:
\bea \label{unitaranq}
U^{\alpha\beta}(t) =
\sum_{\mu\neq\alpha,\beta}|\mu\rangle\langle\mu|
+u^{\alpha\beta}(t),~
\eea
where $u^{\alpha\beta}(t)$ lives in the linear span of $|\alpha\rangle$ and
$|\beta\rangle$ and is unitary. The bounds (\ref{eqEks}) yield for $U^{\alpha\beta}
\Omega U^{\alpha\beta \dagger}$:
\bea \label{evlb}
E_k\geq\Lambda_k=2\left(|P_{\alpha}-P_{\beta}|\left| u^{\alpha \beta}_{1,1}(t)\right|
\hspace{-1mm}\cdot\hspace{-1mm}\left|
u^{\alpha \beta}_{1,2}(t)\right|
\right.
 \nonumber\\
\left.
-\min_{A}\sum_{a\in \Gamma^k_A}\sqrt{P_{\alpha_a}P_{\beta_a}} \right),
\eea
Notice that the right
hand side of (\ref{evlb}) can be made time-independent by
using $\max_t\{|u^{\alpha\beta}_{1,1}(t)|\hspace{-0.7mm}\cdot\hspace{-0.7mm}|
u^{\alpha\beta}_{1,2}(t)|\}=1/2$
-- a consequence of its unitarity. This leads to the desired expression to compute the
$k$-separability of the state under a permutation of diagonal elements.

Finally, note that $E_1$ in (\ref{evlb}), i.e. the first entry of the entropy vector, is equivalent to the PPT criterion. Therefore the bound is exact since the dimension of the subspace is $2x2$ and therefore the PPT criterion is a sufficient and necessary condition. Furthermore, the last bound of (\ref{evlb}), $E_{2^{N-1}-1}$, is also exact as proven in \cite{hashemi}. Therefore, our detection criteria is exact both for entanglement and genuine N-partite entanglement and it can be used to quantify them.  \\


\begin{thebibliography}{99}

\bibitem{nielsen} T.~J.~Osborne and M.~A.~Nielsen, Phys. Rev. A {\bf 66}, 032110 (2002).

\bibitem{amico} L.~Amico, R.~Fazio, A.~Osterloh, and V.~Vedral, Rev. Mod. Phys. {\bf 80}, 517576 (2008).

\bibitem{prange} R. E.~Prange and S. M.~Girvin, {\it The quantum Hall effect} (Springer-Verlag, New York, 1990), 2nd ed.

\bibitem{alicki} R.~Alicki and M.~Fannes, Phys. Rev. E {\bf 87}, 042123 (2013).

\bibitem{ll5} See, for example, L. D.~Landau and E. M.~Lifshitz, {\it Statistical physics, Part I} (Pergamon, New York, 1980).

\bibitem{pusz} W.~Pusz and S.~L.~Woronowicz, Commun. Math. Phys. {\bf 58}, 273 (1978).

\bibitem{lenard} A.~Lenard, J. Stat. Phys. {\bf 19}, 575 (1978).

\bibitem{zhopa} Although $\Delta$ must be $\ll \mathcal{E}_0$, it must be large enough
for $\ln \left(N_\Delta\right)$ to be $\propto N$.

\bibitem{armen} A.~E.~Allahverdyan, R.~Balian, and Th.~M.~Nieuwenhuizen, Europhys. Lett. {\bf 67}, 565 (2004).

\bibitem{parrondo} R.~Marathe and J.~M.~R.~Parrondo, Phys. Rev. Lett. {\bf 104} 245704 (2010).

\bibitem{jarzynski} S.~Vaikuntanathan and C.~Jarzynski, Phys. Rev. E {\bf 83} 061120 (2011).

\bibitem{akepl} A.~E.~Allahverdyan and K.~V.~Hovhannisyan, EPL {\bf 95} 60004 (2011).

\bibitem{ent_perm} L.~Clarisse, S.~Ghosh, S.~Severini, and A.~Sudbery, Phys. Rev. A {\bf 72}, 012314 (2005).

\bibitem{conc} S.~Hill and W.~K.~Wootters, Phys. Rev. Lett. {\bf 78}, 5022 (1997).

\bibitem{ma} Z.-H.~Ma {\it et al.}, Phys. Rev. A {\bf 83}, 062325 (2011).

\bibitem{maju} M.~Huber and J.~I.~de Vicente, Phys. Rev. Lett. {\bf 110}, 030501 (2013).

\bibitem{np} Nevertheless, as for all measures of entanglement, for general states it is NP-hard even to determine whether this measure is nonzero.


\bibitem{l_sep} There can be at least $\lceil N/l\rceil$- and at most $(N-l+1)$-partite entanglement in an $l$-separable state.

\bibitem{n_1} For the ensemble of $N$ spins with gaps $\delta$ it is easy to see that $n_1=\frac{\mathcal{E}_0-\Delta/2}{p_1\delta}$,
where $p_1$ is the excited level population of one spin.

\bibitem{foot} Indeed, the second term in (\ref{evlb}) vanishes because, due to $\mathcal{E}_0\gg\Delta$, at least one of the diagonal
elements under the square root lays outside $[\mathcal{E}_0-\Delta/2,\,\mathcal {E}_0+\Delta/2]$.

\bibitem{cover} T.~M.~Cover and J.~A.~Thomas, ``Elements of Information Theory'' (Wiley, New-York, 1991).

\bibitem{typic} The rude intuition behind this is that the entropy of the probability vector $\{\frac{1}{M}\}_{i=1}^M$ is $\ln M$.

\bibitem{sorensen} O.~W.~S\"{o}rensen, Prog. Nucl. Magn. Reson. Spectrosc. {\bf 21}, 503-569 (1989).

\bibitem{ajm} A.~E.~Allahverdyan, R.~S.~Johal, and G.~Mahler, Phys. Rev. E {\bf 77}, 041118 (2008).

\bibitem{Skr} P.~Skrzypczyk, A.~J.~Short, and S.~Popescu, arXiv:1302.2811 [quant-ph].

\bibitem{LindenSkrzypczyk} N.~Linden, S.~Popescu, and P.~Skrzypczyk, Phys. Rev. Lett. {\bf 105}, 130401 (2010).

\bibitem{brandao} F.~G.~S.~L.~Brand\~ao, M.~Horodecki, J.~Oppenheim, J.~M.~Renes, and R.~W.~Spekkens, arXiv:1111.3882 [quant-ph].

\bibitem{mamaju} M.~Huber, M.~Perarnau-Llobet, J.~I.~de Vicente, Phys. Rev. A {\bf 88}, 042328 (2013).

\end{thebibliography}

\begin{thebibliography}{99}
\bibitem{winter} N.~Linden and A.~Winter, Commun. Math. Phys. {\bf 259}, 129 (2005).
\bibitem{gurvits} L. Gurvits, {\it Classical deterministic complexity of Edmonds' problem and quantum entanglement} in Proceedings of the thirty-fifth annual ACM symposium on Theory of computing, 10 (2003).

\bibitem{maju} M.~Huber and J.~I.~de Vicente, Phys. Rev. Lett. {\bf 110}, 030501 (2013).

\bibitem{mamaju} M. Huber, M. Perarnau-Llobet, J. I. de Vicente, arXiv:1307.3541 [quant-ph].
\bibitem{ma} Z.-H.~Ma {\it et al.}, Phys. Rev. A {\bf 83}, 062325 (2011).

\bibitem{wu} J.-Y. Wu, {\it et al}, 
 Phys. Rev. A {\bf 86}, 022319 (2012).

\bibitem{se} Z.-H. Chen, Z.-H. Ma, J.-L. Chen, S. Severini, Phys. Rev. A {\bf 85}, 062320 (2012).
\bibitem{hashemi} S.~M.~Hashemi Rafsanjani, M.~Huber, C.~J.~Broadbent, and J.~H.~Eberly, Phys. Rev. A {\bf 86}, 062303 (2012).

\end{thebibliography}
\end{document}